# Properties of the Molecular Clouds in NGC 205

L. M. Young

*Astronomy Department, New Mexico State University, Las Cruces, NM 88003*

*Physics Department, New Mexico Institute of Mining and Technology, Socorro, NM 87801*

lyoung@aoc.nrao.edu

## ABSTRACT

The nearby dwarf elliptical galaxy NGC 205 offers a unique opportunity for high resolution studies of the interstellar medium in an elliptical galaxy. The present paper investigates the distribution of molecular gas, molecular line ratios, and the relationships between atomic gas, molecular gas, and dust in this galaxy. The line ratios $^{12}CO(2-1)/(1-0)$ and $^{12}CO(1-0)/^{13}CO(1-0)$ in one of the molecular clouds in NGC 205 are consistent with the ratios found in other elliptical galaxies and in Galactic giant molecular clouds; they suggest that the CO in this cloud is probably subthermally excited. Atomic gas, molecular gas, and dust are very closely associated on scales of $\sim$100 pc; the atomic gas can be understood as photodissociated envelopes around the molecular clouds. The atomic column densities in this galaxy are quite low ($\sim$10$^{20}$ cm$^{-2}$) because the interstellar UV field is relatively low. The total gas-to-dust column density ratios are consistent with Galactic gas-to-dust ratios. In short, the molecular gas in NGC 205 seems to have very similar properties to the familiar molecular clouds in our own Galaxy, except for the low atomic column densities.

*Subject headings:* galaxies: elliptical & lenticular, CD— galaxies:individual(NGC 205)— galaxies:ISM— Local Group— ISM:molecules— ISM:structure

## 1. Introduction

In recent years it has become apparent that elliptical galaxies are not devoid of neutral interstellar gas and dust. Many, if not most, ellipticals show some sign of a neutral interstellar medium (ISM). Nearly half of the ellipticals in a magnitude-limited sample were detected by the IRAS satellite (Knapp et al. 1989; see also the recent analysis by Bregman et al. 1998). Goudfrooij et al. (1994) estimate that some eighty percent of a magnitude-limited sample of ellipticals in the Revised Shapley-Ames catalog contain dust clouds or lanes which would be visible as optical obscuration if the orientation of the dust lane was favorable. Similarly high dust detection rates were found by van Dokkum & Franx (1995) and Ferrari et al. (1999). Fifty to seventy percent of ellipticals also have atomic hydrogen in a quantity $M(HI)/L_B > 10^{-3}$ (Huchtmeier et al. 1995; Knapp et al. 1985; Wardle & Knapp 1986). Statistical work on the molecular gas content of ellipticals has not progressed as far as for atomic gas and dust, but nevertheless, thirty to forty percent of Es or E/SOs which have been observed have been detected (Rupen 1997; Henkel & Wiklind 1997).

If elliptical galaxies contain molecular clouds, they may be forming stars. Therefore, the evolution of an elliptical galaxy depends on the properties of its interstellar medium, especially its molecular gas, and the relationships between the various phases of the ISM. Is atomic gas formed from the molecular gas or *vice versa*, and will the molecular gas be stable on long timescales? Henkel & Wiklind (1997) discuss the atomic and molecular gas in ellipticals with particular emphasis on the relation between the neutral gas and the hot (X-ray emitting) gas and proposed cooling flows. A molecular cloud's properties such as density, cloud size, and magnetic field strength help to determine whether or how the cloud can survive in a hostile bath of hot gas. But the physical properties (e.g. density and temperature) of the molecular gas in ellipticals are poorly known, so these ideas about gas and galaxy evolution cannot yet be evaluated. Indeed, the molecular clouds in elliptical galaxies might be significantly different from Milky Way clouds because, among other things, X-ray heating and heating from stellar velocity dispersions are probably much more important in ellipticals than they are in spirals. The present observations of NGC 205 use molecular line ratios and the distributions of atomic and molecular gas to study the molecular gas's physical properties and its relation to the atomic gas.

The main hindrance to studying the neutral ISM in ellipticals is that elliptical galaxies tend to be rather far away, so the observed fluxes are weak and also the linear resolution is poor. The two major exceptions to this statement are two dwarf elliptical companions to M31, the galaxies NGC 185 and NGC 205. Their distances are less than 1 Mpc. They are the only elliptical galaxies with a neutral interstellar medium that can be observed at resolutions better than 100 pc, which allows us to resolve the individual giant molecular clouds.

It is true that these galaxies are not true ellipticals in some classification schemes; in general, dwarf ellipticals follow a different fundamental plane than giant ellipticals (Ferguson & Binggeli 1994). However, with respect to the factors which might influence the properties of their interstellar media, they can be regarded as ellipticals or as reasonable facsimiles thereof. They have mostly old stellar populations (Lee 1996; Davidge 1992; Martínez-Delgado et al.

1999), so the energetics of the interstellar medium will be very different than in a spiral galaxy with large numbers of young massive stars. They are supported by stellar velocity dispersions (Bender et al. 1991; Held et al. 1992), and they have no spiral density waves. Because NGC 205 has these properties and others in common with "true" ellipticals, the interstellar medium in NGC 205 offers important insights into elliptical galaxies.

The original detection of $^{12}$CO 1-0 emission in NGC 205 was made by Sage & Wrobel (1990), but that work provided no information on the molecular gas distribution or physical properties. Higher resolution observations by Young & Lo (1996, 1997) and Welch, Sage, & Mitchell (1998) found CO emission associated with a dust cloud in the northern part of the galaxy. In fact, the interferometric CO observations of Young & Lo (1996), with a 20 pc beam size, resolved the molecular cloud. Young & Lo and Welch et al. found a few tentative detections of CO emission in other parts of the galaxy, but were unable to study the gas distribution and the relationship between atomic and molecular gas over most of the galaxy. The present paper addresses that issue. Previous work also gave limited information on the molecular gas's physical properties; the present paper improves on that situation with better $^{12}$CO 2-1/1-0, CO/HCN, and CO/HCO+ line ratios, as well as the first $^{12}$CO/$^{13}$CO ratio in NGC 205.

## 2. Observations and data reduction

Observations of the molecular clouds in NGC 205 were made with the IRAM 30m telescope during August 1996. The observed transitions included $^{12}$CO J=1-0 and 2-1, $^{13}$CO J=1-0 and 2-1, HCN J=1-0, and HCO+ J=1-0. Two 3 mm and one 1 mm SIS receivers were used simultaneously. Each receiver was connected to two backends: a filter bank of 256 × 1 MHz channels and an autocorrelator unit giving 449 channels of 312.5 kHz. The telescope was operated in the wobbler switch (nutating subreflector) mode with a cycle time of 4 seconds and a throw of $180''$ in azimuth. The $180''$ throw is similar to the largest diameter of the HI distribution in NGC 205 and may possibly be similar to the north-south extent of the molecular gas in that galaxy. However, the strong velocity gradient of the gas in NGC 205 (Young & Lo 1997) means that gas in the reference position would have a different velocity (by 30 km s$^{-1}$) from gas in the on-source position. This type of confusion is not evident in the data, so the $180''$ throw must have been large enough to avoid problems.

The telescope's focus was checked two or three times every night, especially around the times of sunset and sunrise. The pointing was checked every two to three hours with observations of W3OH; the pointing errors are consistent with the $3.5''$ rms expected for the period soon after a pointing model update (Greve et al. 1996). Pointing scans on Mars also provided checks that the telescope beam size was as expected. Cold load calibrations were made at least every 15 minutes, and more often as the weather degraded. The temperature scale was also checked with periodic observations of W3OH and W51D in all of the transitions mentioned above (Mauersberger et al. 1989) and was found to be accurate and repeatable with a rms of about 13%.

System temperatures ($T_A^*$) were 200–300 K for the HCN, HCO+, and $^{13}$CO lines, and 300–700 K for the $^{12}$CO lines. Integration times were about one hour at each position in the $^{12}$CO lines and 2–3 hours in the HCO+, HCN, and $^{13}$CO lines. After eliminating bad scans, the data were averaged together using a standard weighting by integration time and system temperature. Baselines of order 0 or 1 were subtracted from the spectra. All data presented in this paper are in the main beam brightness temperature scale, which is obtained from the antenna temperatures via the relation $T_{mb} = T_A^*/\eta_{mb}$. The beam sizes, velocity resolutions (the width of a 1 MHz channel), and the values of $\eta_{mb}$ (the ratio of the beam to forward efficiencies) are given in Table 1.

Table 2 gives the coordinates of the positions observed in NGC 205. The positions were mostly selected based on the presence of dust, which is visible as obscured patches in optical images (Hodge 1973, Peletier 1993, Kim & Lee 1998). Hodge (1973) assigned numbers to the dust clouds he detected, and that nomenclature is followed in this paper.

Tables 3 and 4 give the results of Gaussian fits and moment analysis for each position and each line observed. The final stacked, baseline-subtracted spectra were fit with one Gaussian component if there appeared to be a detection. For the fits that converged, the table gives the component area, central velocity, and width (full width at half maximum) along with their associated formal uncertainties.

Tables 3 and 4 also give the line area derived from a simple integration over velocity. After integrating over velocity ranges of 10, 20, 30, 40, and 50 km s$^{-1}$, it was found that velocity ranges of 30 km s$^{-1}$ worked well for all the detected lines except those in DC2, where 40 km s$^{-1}$ is necessary. For clouds with no detected emission, a broader velocity range of 50 km s$^{-1}$ is used. The velocity ranges were all centered on the velocity of the Gaussian fit component of $^{12}$CO(1-0) in that cloud, if there was a reliable fit, or $-235.6$ km s$^{-1}$ if there wasn't. The $1\sigma$ uncertainty in the integrated line area counts two contributions (summed in quadrature): the error in determining the baseline level and the error from summing individual noisy channels (e.g. Sage & Isbell 1991). In the present case, the primary contribution to uncertainty in integrated area is from summing the noise over the line. Of course this estimate is only a formal uncertainty; it underestimates the true uncertainty because it ignores correlations from one channel to the next (baseline wiggles). Despite this drawback, the uncertainties in the integrated areas are probably more reliable than the formal errors to the Gaussian fits. The $1\sigma$ values in Tables 3 and 4 do not include the uncertainty in the temperature calibration scale, which is probably about 10–15% (see above).

## 3. Molecular gas distribution and kinematics

Figure 1 shows the observed positions with respect to the atomic hydrogen (HI) column density and the dust clouds in NGC 205. The HI distribution was mapped with the Very Large Array[1] and was previously published by Young & Lo (1996, 1997). Figures 2, 3, 4, and 5 show $^{12}$CO(1-0) spectra and HI spectra at the positions with detected CO; in figures 2 and 4, the spectra are laid out in their relative positions on the sky. CO emission is detected from three clouds: Hodge's (1973) dust cloud number 11, which is called DC11 in this paper, and is about $1'$ northeast of the galaxy center; DC2, near the galaxy center, and DC12, about $1.5'$ southeast of the galaxy center. The emission in DC11 was already known (Young & Lo 1996, 1997; Welch et al. 1998). Young & Lo (1996) also showed one tentative detection in the southern cloud DC12, whereas the present observations detect CO in several positions within DC12. The new data also show detected emission from DC2, near the center of the galaxy, for the first time.

All positions except for the ones called "NoDust1" and "NoDust2" were selected because of dust patches visible in the optical images (see especially Kim & Lee 1998). The positions "NoDust1" and "NoDust2" were selected as an attempt to test whether the molecular gas distribution makes a complete ring around the center of the galaxy, but no gas was detected there. Thus, the atomic gas in NGC 205 clearly follows the distribution of dust patches. The pattern of CO detections and non-detections suggests that the molecular gas also follows the dust distribution, which would be consistent with the results of Sage & Galletta (1993) for three other peculiar ellipticals. However, it should be noted that a complete map of molecular gas in NGC 205 has not yet been made.

Figures 2, 3, 4, and 5 also show the kinematic relationship between the atomic and molecular gas. Both species show the same velocity gradient of about 30 km s$^{-1}$ from the north end of the galaxy to the south end. The HI and CO lines have approximately the same width (FWHM about 10 km s$^{-1}$); in some cases the HI is broader than the CO, but never narrower. The two species have velocity centroids which agree to within 5 km s$^{-1}$ in most cases; the major exception to this statement is the center position of DC12, where the line

---

centroids differ by almost 10 km s$^{-1}$.

The close spatial and kinematic association of atomic and molecular gas in this galaxy suggests that the two phases are probably physically associated with each other. A high resolution map of CO emission in DC11, made with the BIMA millimeter interferometer, strengthens this impression (Young & Lo 1996). I suggest that the atomic gas forms a photodissociated envelope around the molecular clouds. The velocity offsets of a few km s$^{-1}$ between the atomic and molecular phases should not be interpreted as evidence against the photodissociation model. Indeed, detailed observations of the Orion molecular cloud–photodissociation region clearly show that the various molecular and atomic species have velocities which differ by several km s$^{-1}$. In Orion A, the $^{12}$CO 7-6 line and the [CII] 158$\mu$m line differ by 2–3 km s$^{-1}$, and the ionized gas differs by about 14 km s$^{-1}$ from the neutral phases (Genzel & Stutzki 1989; see especially their Table 1 and Figure 5). Thus the atomic and molecular gas in NGC 205 could still be physically associated, despite velocity offsets of a few km s$^{-1}$. The photodissociation model of molecular and atomic gas in NGC 205 is dicussed further in section 6.

## 4. H$_2$ masses

Table 3 gives the mass of molecular gas at each position where the integrated intensity is greater than three times its own uncertainty. The $^{12}$CO(1-0) integrated intensity I$_{CO}$ (K km s$^{-1}$, in the brightness temperature scale) is converted to a H$_2$ column density N(H$_2$) = X · I$_{CO}$ mol cm$^{-2}$. The CO-to-H$_2$ conversion factor X is taken to be 1.56×10$^{20}$ mol cm$^{-2}$ per K km s$^{-1}$, which is the value recently determined for our own Galaxy from models of gamma ray emission (Hunter et al. 1997). Integrated CO intensities are also converted to H$_2$ mass per beam

$$M(H_2) = 4.27 \times 10^{-19} \, X \, \Theta^2 \, D^2 \, I_{CO} \quad M_\odot,$$

where the beam size $\Theta$ is in arcseconds and the distance D is in Mpc. The distance to NGC 205 is assumed to be 0.85± 0.1 Mpc (Saha, Hoessel, & Krist 1992; see also Mould, Kristian, & Da Costa 1984, and Salaris & Cassisi 1998). The beam size of the 30m telescope in the $^{12}$CO(1-0) line is 21$''$, or 87 pc at 0.85 Mpc. It is not clear that the CO-to-H$_2$ conversion factor should be the same in NGC 205 as in our own Galaxy. In fact, the photodissociation region models of Kaufman et al. (1999) imply that the CO-to-H$_2$ conversion factor for a molecular cloud could vary by orders of magnitude in different kinds of conditions. However, in this case the Galactic value is used in lieu of something better.

When describing the molecular gas in our own Galaxy, Blitz & Williams (1999) discuss a spectrum of cloud sizes. At the massive end of the spectrum are the self-gravitating giant molecular clouds (GMCs) which have masses $\geq 10^4$ $M_\odot$; at the low-mass end are the non-self-gravitating "chaff" clouds ($10^2$ $M_\odot$). From Table 3 it is apparent that the molecular clouds in NGC 205 have masses $\geq 10^4$ $M_\odot$. Furthermore, observations of $^{12}CO(1-0)$ emission with the Berkeley-Illinois-Maryland interferometer (BIMA) resolved the molecular gas in the cloud DC11 and show that its largest detectable extent is about $30'' = 120$ pc (Young & Lo 1996). Thus, the molecular clouds in NGC 205 have similar sizes and masses to the large Galactic GMCs, not the "chaff" clouds.

The $H_2$ masses in Table 3 agree well with previous measurements. For example, the $H_2$ mass detected in the center position of dust cloud 11 is given as $(5.7\pm0.3)\times10^4$ $M_\odot$. (The uncertainty quoted here includes just the measurement noise, not uncertainties in the calibration scale.) The mass given by Welch et al. (1998) from NRAO 12m observations of the same cloud is $(6.0\pm1.7)\times10^4$ $M_\odot$ in a larger beam, after correction to the same CO-to-$H_2$ conversion factor used here. (Welch et al. used a conversion factor of $2.3\times10^{20}$ mol cm$^{-2}$ per K km s$^{-1}$ [Strong et al. 1988].) The mass of this cloud from the interferometric CO map of Young & Lo (1997) is $(6.4\pm1.3)\times10^4$ $M_\odot$. Thus, there is good consistency between the detected fluxes of indiviual clouds.

The total $H_2$ mass for NGC 205 is more difficult to constrain because no observation has covered the entire galaxy and because the positions which have been observed overlap each other. A sum over the firm ($> 3\sigma$) detections, counting only the non-overlapping positions, gives $(9.9\pm0.5)\times10^4$ $M_\odot$, which may be an underestimate to the total. Of this sum, the bulk of the gas (60%) is in the northern cloud DC11. But $2.2\times10^4$ $M_\odot$ comes from positions in the southern half of the galaxy, in DC12, where Welch et al. (1998) did not find any CO emission.

With a molecular gas mass of at least $9.9\times10^4$ $M_\odot$, and an HI mass of $(4.3\pm0.6)\times10^5$ $M_\odot$ (Young & Lo 1997), the $H_2$/HI mass ratio in NGC 205 is $\geq 0.23$, which is consistent with the mass ratios found in other ellipticals. The FIR-selected samples of Wiklind et al. (1995) and Huchtmeier et al. (1995) have $H_2$/HI mass ratios ranging from about 0.03 to 1 or 2.

## 5. Line ratios

Figures 6 and 7 compare spectra of the different molecular species which were observed in NGC 205. Table 5 gives the line ratios that could be measured with the current dataset, estimated both from Gaussian fits and from integrated line areas (on the main beam brightness temperature scale). Where the weaker line did not have a reliable Gaussian fit, no Gaussian line ratio is given, but an integrated area ratio is given. If the weaker line has an integrated area with less than $3\sigma$ significance, the ratio is a lower limit. In those cases the $3\sigma$ limit is used for the weaker line so that the lower limits are always conservatively low.

For the ratio of two lines in the same observing band (1-0/1-0 or 2-1/2-1), no correction was made for the difference in beam sizes. For the 1-0/2-1 ratios, the 2-1 data were convolved to the same spatial resolution as the 1-0 data. To make this convolution, dust clouds 11 and 12 were mapped in a five-point cross pattern spaced at $10''$ (approximately the FWHM of the 2-1 beam). To the extent that the beams can be approximated by Gaussians of the FWHM given in Table 1, the $^{12}$CO(2-1) data need to be convolved with an $18''$ Gaussian function to bring them to the same resolution as the 1-0 data. If this convolution is changed into a discrete sum at $10''$ steps, the relative weights are 1.0 for the center pointing and 0.426 for each of the offset pointings. (If data had been taken on a nine-point square instead of a five-point cross, the relative weights of the corner points would be 0.181, so that the convolution is still dominated by the center point and the four nearest points.) To conserve the detected flux, the convolution is normalized by the sum of the weights.

Thus, the convolution of the $^{12}$CO(2-1) data with an $18''$ Gaussian function can be approximated by the sum

$$0.370 \left( S(0'') + 0.426 \sum S(10'') \right),$$

where $S(0'')$ represents the spectrum at the center position and $\sum S(10'')$ is the sum of the four offset spectra. When comparing $^{13}$CO(2-1) to $^{13}$CO(1-0), the weights are 0.459 for the offset pointings and 0.352 for the normalization. This formula is very close to the one used by Allen et al. (1995). It is significantly different from the one used by Johansson et al. (1994), who assumed that the response of the 1-0 beam is constant over the area of the 2-1 beam. The present paper makes no attempt to correct for the fact that a molecular cloud may have different sizes in the different molecules and different transitions.

The $^{12}$CO(1-0)/(2-1) line ratios for DC11 in Table 5, namely $2.2\pm0.3$ and $1.9\pm0.2$, give 2-1/1-0 ratios of about $0.5\pm0.1$. This result is significantly different from the value of 0.9 $\pm0.2$ previously reported by Young & Lo (1996). The new data are higher quality and the lines are detected with better signal-to-noise ratios, so the value of $0.5\pm0.1$ should be adopted. This ratio indicates subthermal excitation of the CO transitions, which may arise from either low kinetic temperatures and/or low densities. The corresponding line ratio in the southern cloud, DC12, is consistent with the value found in DC11. Also, the present $^{12}$CO(1-0)/HCN(1-0) limit in NGC 205 is $^{12}$CO/HCN $> 13$; that is a significant improvement over the previous work of Welch et al. (1998), who found $^{12}$CO/HCN $> 6$.

Table 6 compares the line ratios measured in the cloud DC11 of NGC 205 to ratios measured in other elliptical galaxies, in spiral galaxy disks, and in spiral galaxy nuclei. The table is not intended to be a complete review of the literature, merely to provide some interesting comparisons. The values for NGC 205 are taken from Table 5, after averaging the results from the integrated areas and the Gaussian fitting. Of course, these two methods do not give independent results and the uncertainty in the ratio is not assumed to decrease by that averaging. The error estimates for NGC 205 in Table 6 now *include* the uncertainty in the calibration scale.

The available line ratios indicate that the molecular clouds in NGC 205 have broadly similar physical properties to the moderate density giant molecular clouds in our own spiral's disk, though not to the dense molecular gas in the nuclei of galaxies. The density information comes from the high density tracers HCO+ and HCN; the lower limits in NGC 205 are consistent with the high values found in spiral galaxy *disks*, but not with the low values found in the *nuclei* of spirals or the circumnuclear disk of Cen A. The $^{12}CO/^{13}CO$ ratio in NGC 205 is consistent with all the other $^{13}CO$ ratios in the table.

Finally, the $^{12}CO$ 2-1/1-0 ratio in NGC 205 is consistent with the values measured in other ellipticals and in the Milky Way disk, but not with the very low values from the center of M31. Often this line ratio is used as a very rough indicator of gas temperature; for example, the $^{12}CO$ 2-1/1-0 values $\geq 1$ which are found in the Small Magellanic Cloud (Rubio et al. 1993a, 1993b; Lequeux et al. 1994) are thought to arise in relatively warm gas bathed in a high UV and cosmic ray field. The values of 0.2–0.4 which are found in the center of M31 (Allen et al. 1995) are thought to arise in very cold gas with a temperature of about 5 K. The simple models of Braine & Combes (1992) and Allen et al. (1995) would imply that temperatures less than about 7 K are required for the gas in NGC 205. However, without a more complete analysis it should not be assumed that these models are appropriate for dust clouds 11 and 12 in NGC 205.

## 6. Molecular and atomic gas

Table 7 gives the atomic, molecular, and total gas column densities for all the positions observed with the 30m telescope. The most striking result from the table is the low atomic column densities which are associated with the molecular gas. The highest atomic column density in the table (also the highest atomic column density in the galaxy, at $21''$ resolution) is $4\times10^{20}$ cm$^{-2}$, and the strongest CO emission in the galaxy comes from a position with an HI column density of just $2\times10^{20}$ cm$^{-2}$.

This result is surprising because it runs counter to the conventional wisdom that molecular gas cannot form unless HI column densities are about $10^{21}$ cm$^{-2}$ or greater. For example,

Blitz (1993) analyzed the relationship between Galactic giant molecular clouds (GMCs) and HI, and found that the Galactic GMCs are associated with atomic clouds of peak column density 1–2×10$^{21}$ cm$^{-2}$. In ultraviolet absorption experiments towards Galactic OB associations, Savage et al. (1977) found that molecular hydrogen usually does not form until the HI column density reaches 5×10$^{20}$ cm$^{-2}$. A similar result is found by Lada et al. (1988) for a spiral arm in the galaxy M31; in that case, a linear fit to molecular and atomic column densities implies that molecular gas does not form until the atomic column density is at least 10$^{21}$ cm$^{-2}$. Furthermore, in the observations of Lada et al., molecular gas (CO emission) is not *detectable* until the atomic hydrogen column density rises to about 5×10$^{21}$ cm$^{-2}$. All of this evidence is interpreted to mean that a layer of atomic hydrogen of about 10$^{21}$ cm$^{-2}$ must be present to shield molecular hydrogen from dissociation by the interstellar UV field. Yet molecular hydrogen has clearly formed in NGC 205 in locations with atomic column densities of 2×10$^{20}$ cm$^{-2}$ and lower.

The comparison between NGC 205 and the M31 results of Lada et al. (1988) is particularly appropriate because NGC 205 and M31 are at comparable distances (NGC 205 is a companion of M31) and the two galaxies have been studied with the same observing techniques at comparable spatial and linear resolutions. Figure 8 makes a graphical comparison between HI and H$_2$ column densities in M31 and NGC 205. The data for M31 have been corrected to the same CO/H$_2$ conversion factor which is adopted in this paper (Section 4) and have been corrected for an inclination of 77°. The figure shows that the molecular clouds studied by Lada et al. in M31 have similar H$_2$ column densities to the clouds in NGC 205. (More specifically, the brightest cloud in NGC 205, DC11, is similar to the median clouds in Lada's sample.) But the clouds in NGC 205 have smaller atomic column densities, by about a factor of six. One cannot conclude that the molecular clouds in NGC 205 are different from *all* of the clouds in M31 or in our Galaxy, but they certainly *are* different from the "typical" or most commonly studied GMCs in M31 and in our own Galaxy.

The photodissociation models of Draine & Bertoldi (1996) provide an explanation for the threshold effect seen in the formation of molecular gas in Galactic (and M31) GMCs, and they also explain why the molecular clouds in NGC 205 are different. Molecular hydrogen can exist on long timescales if its formation rate balances the photodissociation rate. Thus the relative amounts of atomic and molecular gas are strongly dependent on the ratio $\chi/n_H$, where $\chi$ measures the intensity of the UV radiation field (at 1000 Å) in units of the Habing (1968) value, and $n_H$ is the gas volume density. When $\chi/n_H$ is high— 0.1 to 10, as in the case of GMCs which are associated with young massive stars, the models of Draine & Bertoldi (1996) show that the formation of molecular gas does indeed require a shielding atomic layer of about 10$^{21}$ cm$^{-2}$. Furthermore, the formation of molecular gas happens very quickly, in a small range of values around 10$^{21}$ cm$^{-2}$. However, when $\chi/n_H$ is low, either due to a low UV field or a high gas density, molecular gas can form with much smaller shielding atomic layers; the shielding layer can be $10^{18}$ cm$^{-2}$ or even lower. The molecular and atomic column densities observed in M31 by Lada et al. (1988) are consistent with $\chi/n_H \sim 0.1$, but the results in NGC 205 are more consistent with $\chi/n_H \sim 0.01$ (see Draine & Bertoldi's figures 9 and 10).

The fundamental difference between the molecular clouds in NGC 205 and those in M31 is probably not a higher density $n_H$, but a lower UV field $\chi$. That is because the high density molecular tracers have not been detected in NGC 205. Also, Welch et al. (1998) have attempted to estimate the strength of the interstellar UV field at the positions of the molecular clouds in NGC 205 by counting the known bright blue stars and estimating the UV field produced by the old stellar population. They estimate that the UV field in NGC 205 is about one tenth as strong as the standard solar neighborhood field. Thus, I speculate that a lower UV field in NGC 205 makes it easier to form and retain molecular gas in that galaxy than in the disk of our own Galaxy.

When the atomic gas in ellipticals can be mapped, the peak HI column densities are often found to be small, around $10^{20}$ cm$^{-2}$, which leads authors to suggest that molecular gas cannot form and therefore star formation cannot occur (e.g. Oosterloo, Morganti, & Sadler 1999). These assumptions are based on comparisons with Galactic GMCs; however, the relationship between atomic and molecular gas may be different in ellipticals than it is in our Galaxy. A proper understanding of molecular gas and star formation in ellipticals must await future surveys of molecular gas, rather than inferences based on the atomic gas.

## 7. Gas and dust

In elliptical galaxies, the gas-to-dust ratio may be an important clue to the evolution of the ISM. For example, dust which is acquired in a merger with a gas-rich galaxy may be gradually destroyed by the hot gas in the elliptical (e.g. Henkel & Wiklind 1997). But a simple comparison of dust opacities and gas column densities in NGC 205 indicates that the gas-to-dust ratio in this galaxy is probably consistent with the Galactic value. Table 8 gives the atomic plus molecular column densities in dust clouds 2 and 11, taken from Table 7, and some estimates of the dust opacity in those clouds.

The first step in the estimation of a gas-to-dust ratio is to correct the observed dust opacities for the fact that the dust clouds are probably behind some of the stars in the galaxy. Line 2 of Table 8 gives the peak extinctions $A_V$, in magnitudes, measured by Price & Grasdalen (1983) [PG83] for two clouds in NGC 205. One may correct these observed extinctions to the true extinctions by assuming that the dust is in a small screen about halfway through the galaxy (Line 3; also taken from PG83). Line 4 gives the $N_H/A_V$ values derived from this information. For comparison, the average Galactic relations $N_H/E(B-V) = 5.8 \times 10^{21}$ cm$^{-2}$ mag$^{-1}$ (Bohlin, Savage, & Drake 1978) and $A_V/E(B-V) = 3.1$ (Clayton, Cardelli, & Mathis 1989) give $N_H/A_V = 1.9 \times 10^{21}$ cm$^{-2}$ mag$^{-1}$.

The $N_H/A_V$ estimates for DC11 are similar to this canonical Galactic value. The estimates for dust cloud 2 are factors of 10–40 lower than the canonical Galactic value, which is probably related to the poor resolution of the gas column density measurements. Figure 1 shows that DC2 is significantly smaller than the 21″ beam used to measure gas column densities. The $A_V$ measurements are the peak values in the cloud at optical resolutions, but the gas column densities are smoothed to resolutions of 21″. DC11 is well matched to the size of the 21″ beam, but the poor radio resolution probably underestimates the gas-to-dust ratio there as well.

The true visual extinctions are themselves uncertain because of the unknown distributions of gas and stars. For example, Witt & Gordon (1999) make a model in which dust is uniformly mixed with stars in the inner third of an elliptical galaxy; they include the effects of scattering, and they predict true visual extinctions which are about a factor of five larger than those of PG83 for the same observed extinction. Thus, the gas-to-dust ratios measured here are highly uncertain. However, the data are consistent with the interpretation that the clouds in NGC 205 have the same gas-to-dust ratio as Galactic clouds. Welch et al. (1998) have also compared the dust mass in NGC 205 (measured from FIR emission) to the gas mass, and they find gas-to-dust ratios somewhat lower than, but roughly comparable to, the canonical Milky Way ratio of ∼100. If the gas-to-dust ratio can indeed be used to tell us something about the destruction of dust in ellipticals (as suggested above), then these broad consistencies imply that the environment inside of NGC 205 is not hostile enough to destroy much dust or, possibly, that the gas and dust in NGC 205 were acquired relatively recently.

## 8. Summary

The molecular clouds in NGC 205 have properties which are similar to the giant molecular clouds in the disk of our own Galaxy. For example, gas-to-dust ratios have been estimated from optical extinctions and atomic and molecular gas column densities. These ratios are highly uncertain because of the different spatial resolutions of the optical and radio data, and because of the unknown geometry of the dust with respect to the stars. However, the data are consistent with the interpretation that the gas-to-dust ratio in the molecular clouds of NGC 205 is similar to the gas-to-dust ratio in our own Galaxy (section 7). The molecular line ratios $^{12}CO(2-1)/(1-0)$, $^{13}CO(1-0)/^{12}CO(1-0)$, and HCN and HCO+ upper limits tell a similar story: the line ratios in NGC 205 are similar to those in giant molecular clouds in the disk of our own Galaxy (section 5).

However, a detailed analysis of the relationship between atomic and molecular gas shows that in this respect, the molecular clouds in NGC 205 are significantly different from their Galactic counterparts. The associated HI envelopes in NGC 205 have much smaller column densities than the HI envelopes around Galactic GMCs. The idea that $10^{21}$ atoms cm$^{-2}$ of atomic gas is necessary for the formation of molecular gas is a rule of thumb which does happen to work (most of the time) for Galactic GMCs and also for molecular clouds in the disk of M31 (section 6). However, this is not an intrinsic property of molecular clouds; it depends on the local conditions such as the gas density and the interstellar UV field. In the dwarf elliptical NGC 205, molecular gas forms with smaller atomic envelopes, which is probably because the interstellar UV field is lower than in the solar neighborhood. In general, because of this dependence on the UV field, it may be easier to form and retain molecular gas in low-luminosity elliptical galaxies than in the spirals such as our own. But the molecular clouds which do form are expected to have properties which are similar to spiral galaxy molecular clouds, and the elliptical galaxies may be forming stars just as the spirals do.

Thanks to G. Welch, M. Rupen, B. Draine, A. Witt, and R. Young Owl for discussions about elliptical galaxies and the ISM. The referee, L. J. Sage, provided many helpful suggestions. The majority of this paper was completed while L.Y. was a Tombaugh Scholar at New Mexico State University. Thanks also to the Physics Department at New Mexico Tech for their hospitality.

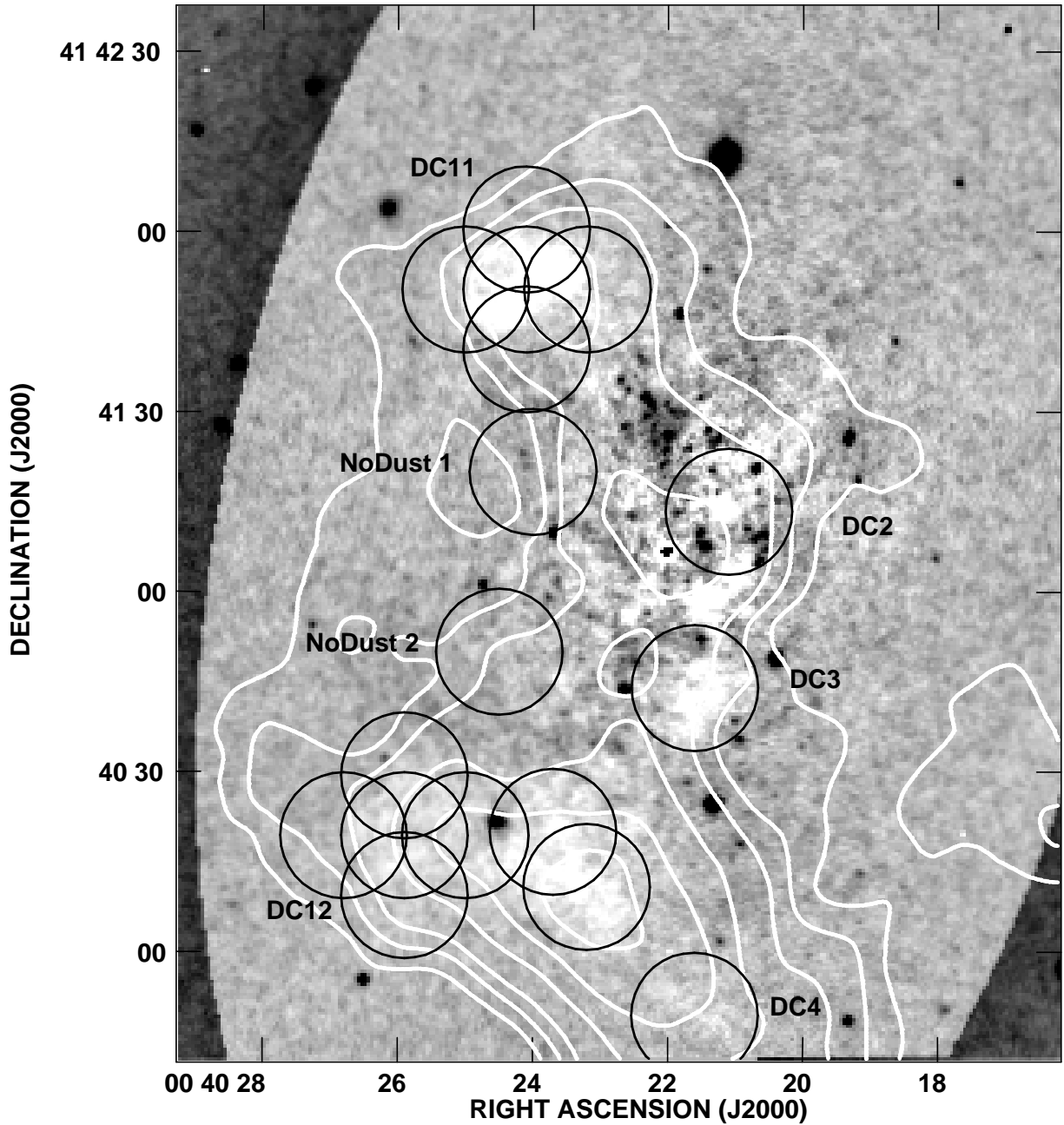

Fig. 1.— Locations of CO pointings with respect to atomic gas and dust in NGC 205. The greyscale is a broadband B image of NGC 205, courtesy of R. Peletier (see Peletier 1993). Smooth elliptical isophotes were fit to the image (after masking bright stars or clusters and dust clouds), and the smooth model was subtracted. Regions of dust obscuration show up as white patches and are labelled after the nomenclature of Hodge (1973); see also Kim & Lee (1998). White contours show the distribution of atomic hydrogen (see Young & Lo 1997) at 21″ resolution; contour levels are 10, 20, 30, 50, 70, and 90% of the peak $(4.5 \times 10^{20}$ cm$^{-2})$. Black circles show the pointings observed at the IRAM 30m telescope and the 21″ FWHM beam size of the $^{12}$CO(1-0) data.

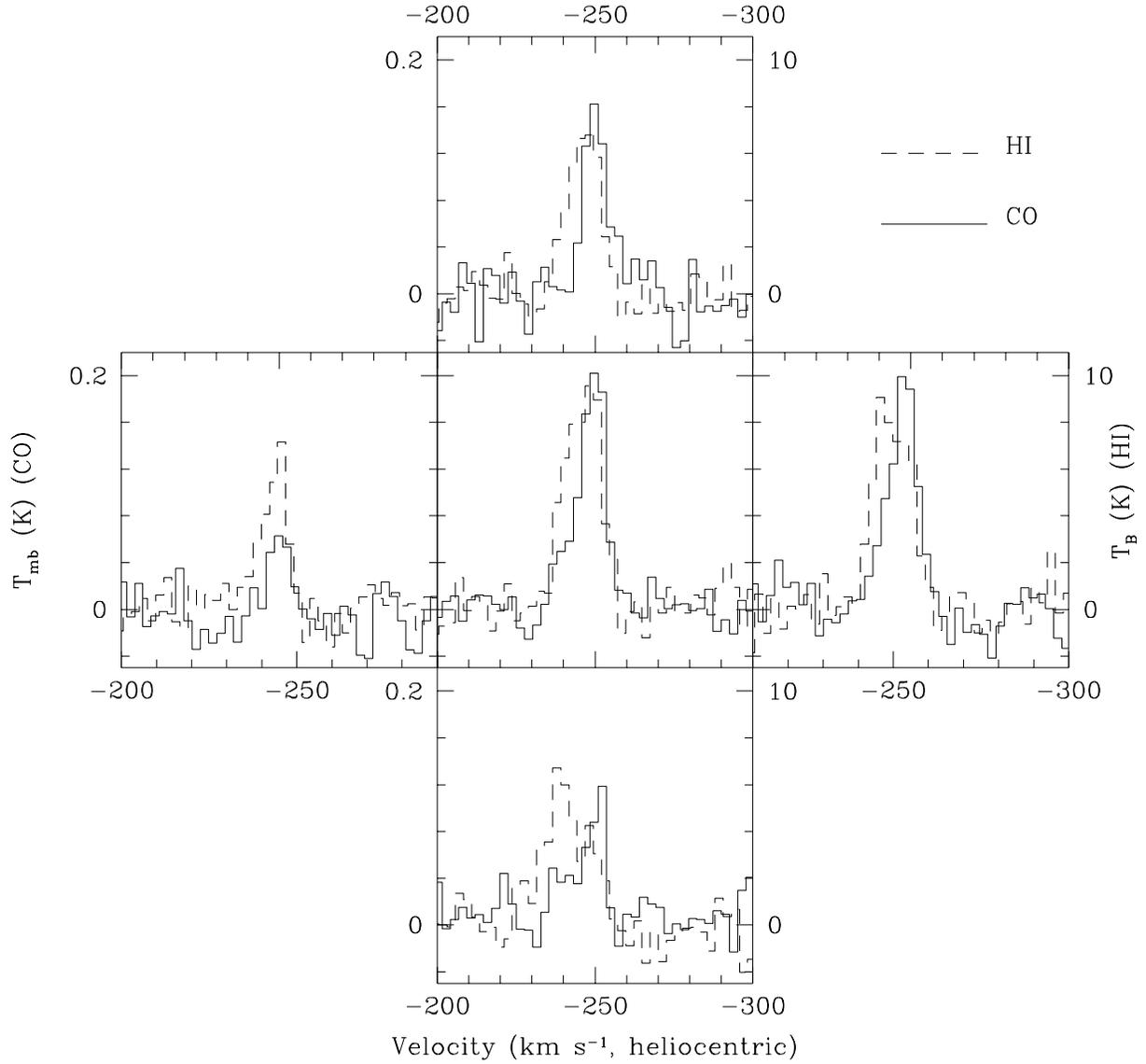

Fig. 2.— Spectra of $^{12}$CO(1-0) and HI in dust cloud DC11. Each panel is a different pointing separated by 10″, and the panels are laid out in their respective positions on the sky (see figure 1). The intensity scale on the left is for CO and the scale on the right is for HI. The molecular spectra in this and subsequent figures cover a larger velocity range than what is shown.

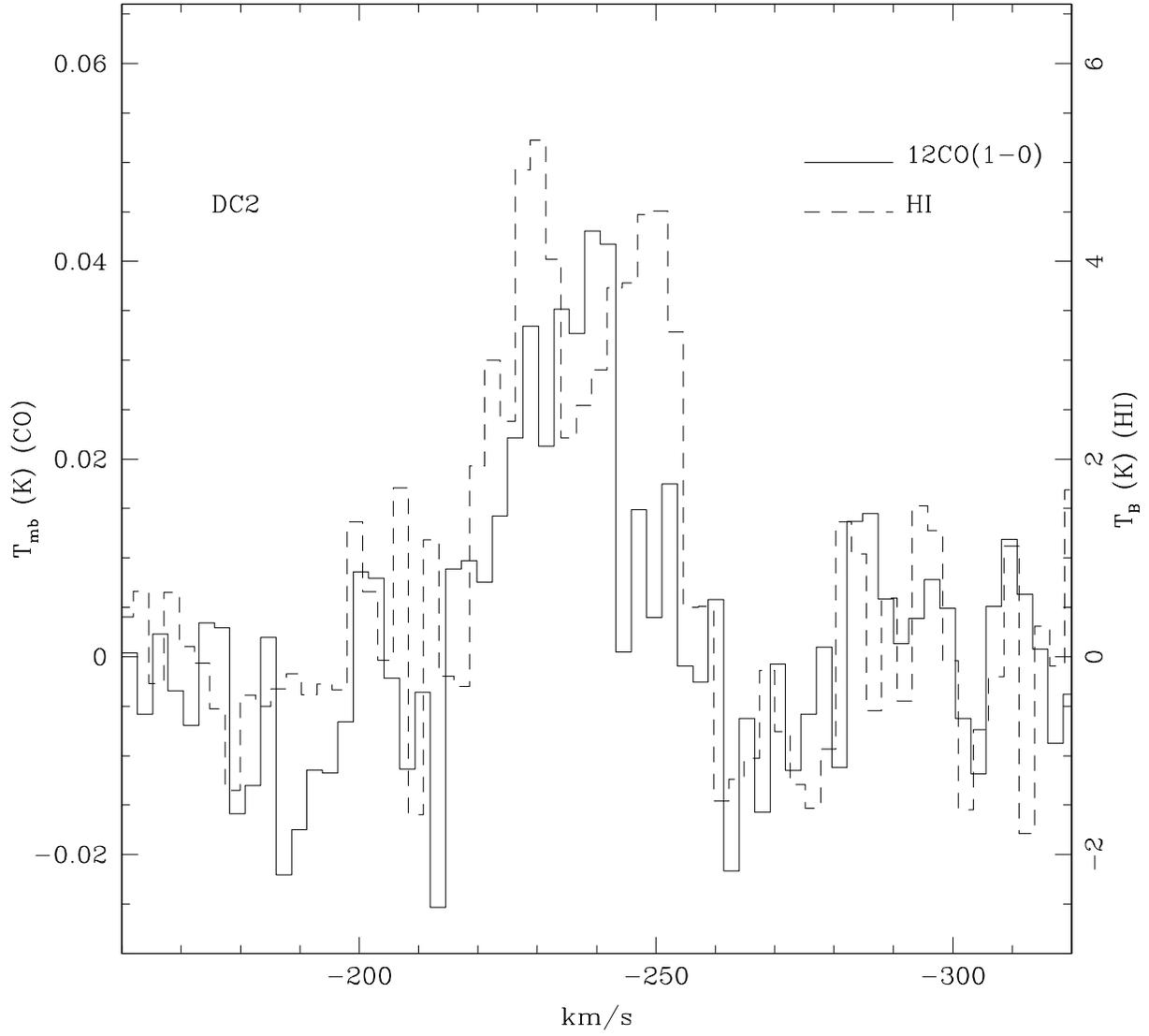

Fig. 3.— Spectra of $^{12}$CO(1-0) and HI in DC2 (a small, dark cloud near the center of the galaxy).

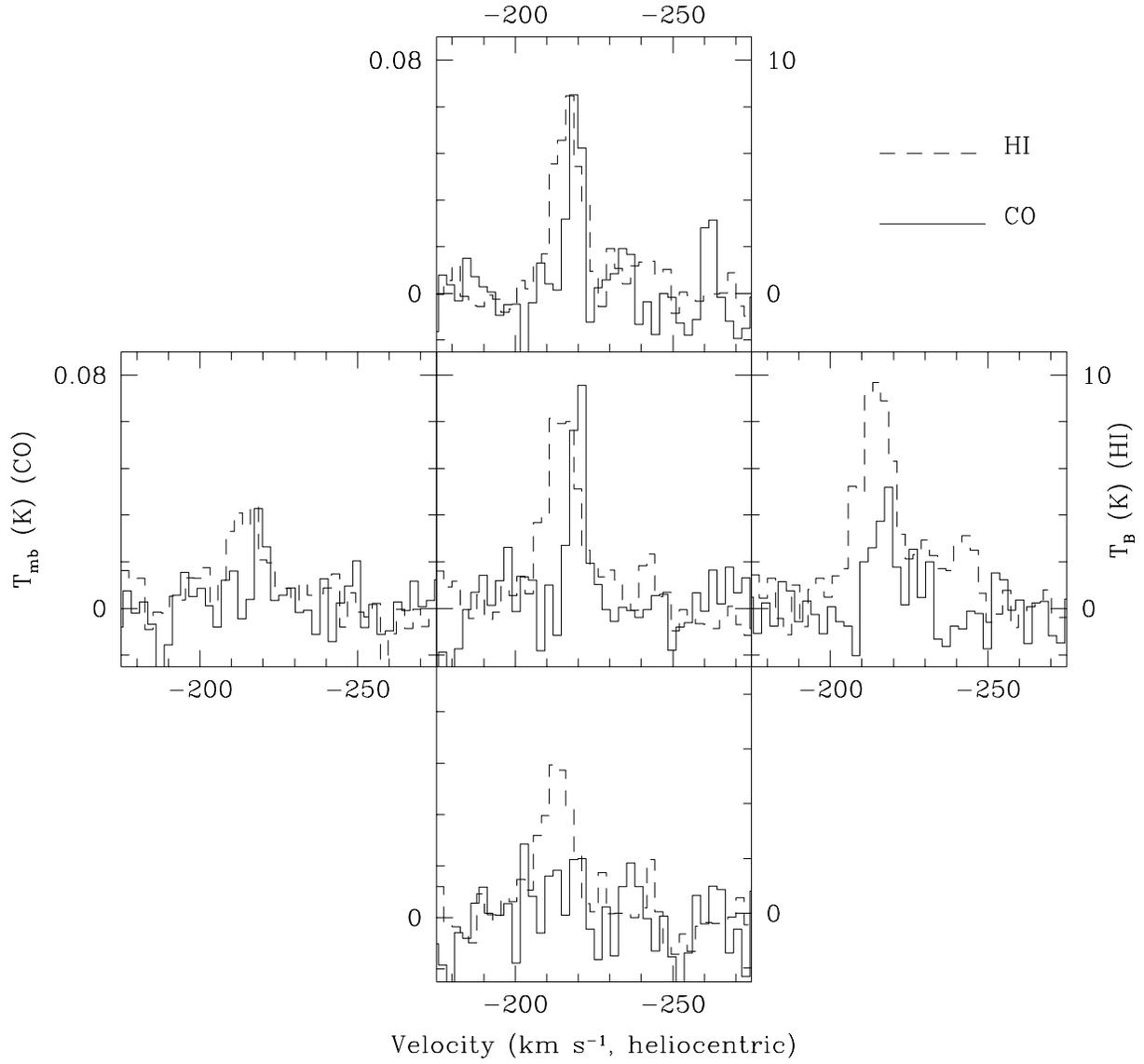

Fig. 4.— Spectra of $^{12}$CO(1-0) and HI in the eastern part of cloud DC12 (a large, irregularly shaped cloud in the southern part of NGC 205; see figure 1).

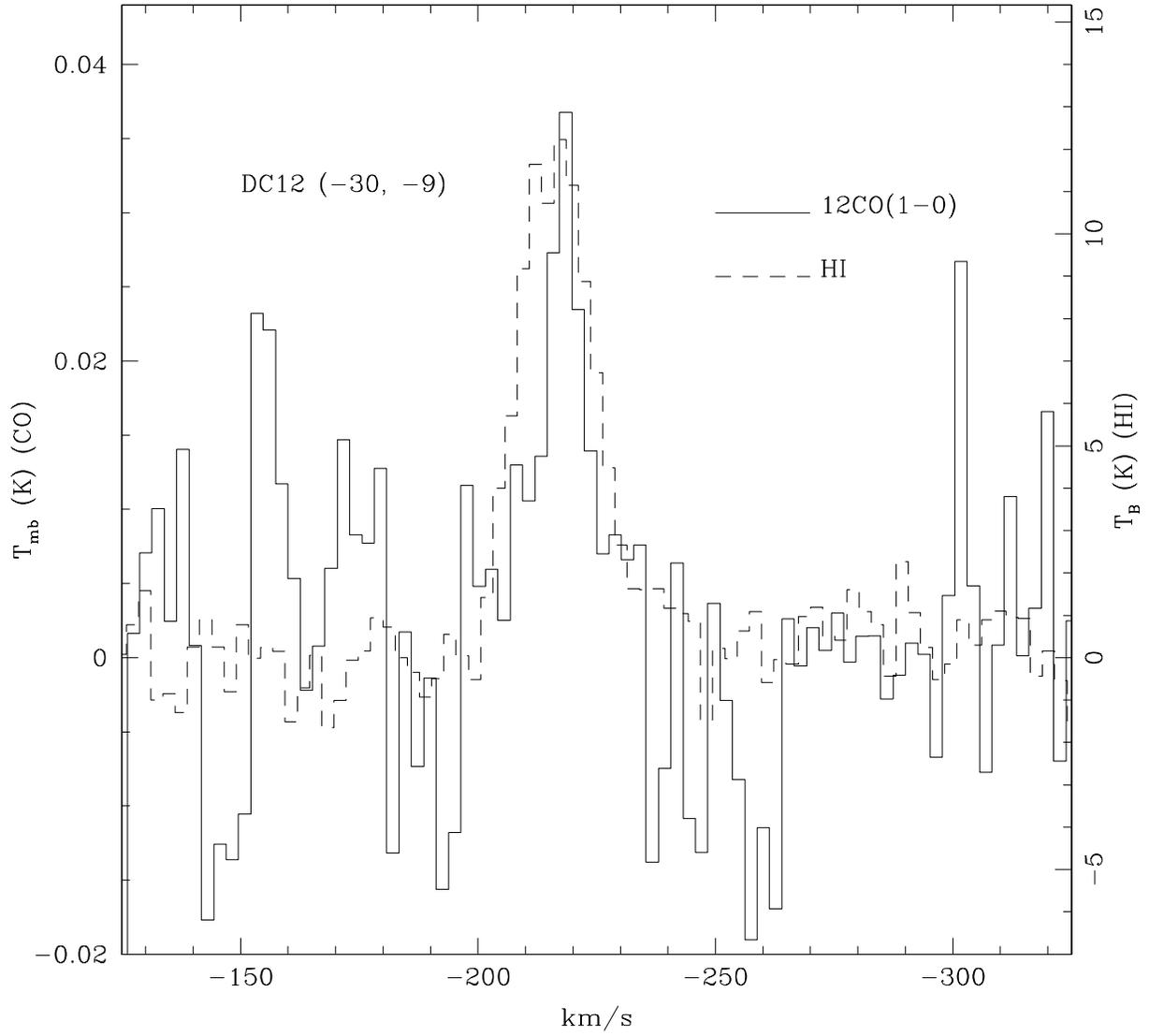

Fig. 5.— Spectra of $^{12}$CO(1-0) and HI in the southwestern extension of cloud DC12. This location has the highest HI column density in the galaxy.

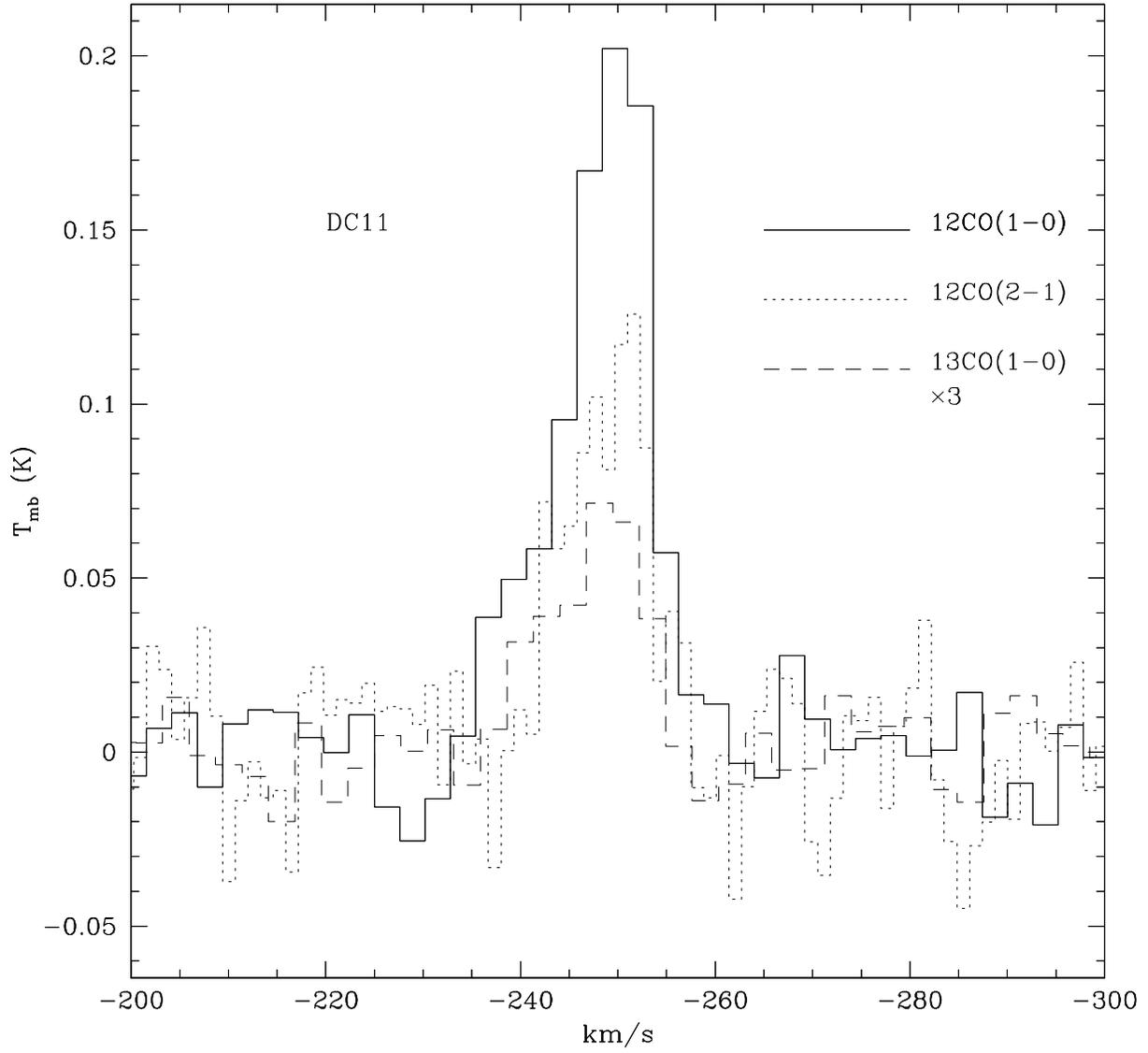

Fig. 6.— Spectra of $^{12}$CO(1-0), $^{12}$CO(2-1), and $^{13}$CO(1-0) in DC11. $^{12}$CO(1-0) and $^{12}$CO(2-1) are shown at the same intensity scale, but $^{13}$CO(1-0) is scaled up for visibility. The $^{12}$CO(2-1) data have been convolved to the same spatial resolution as the $^{12}$CO(1-0) data (see text).

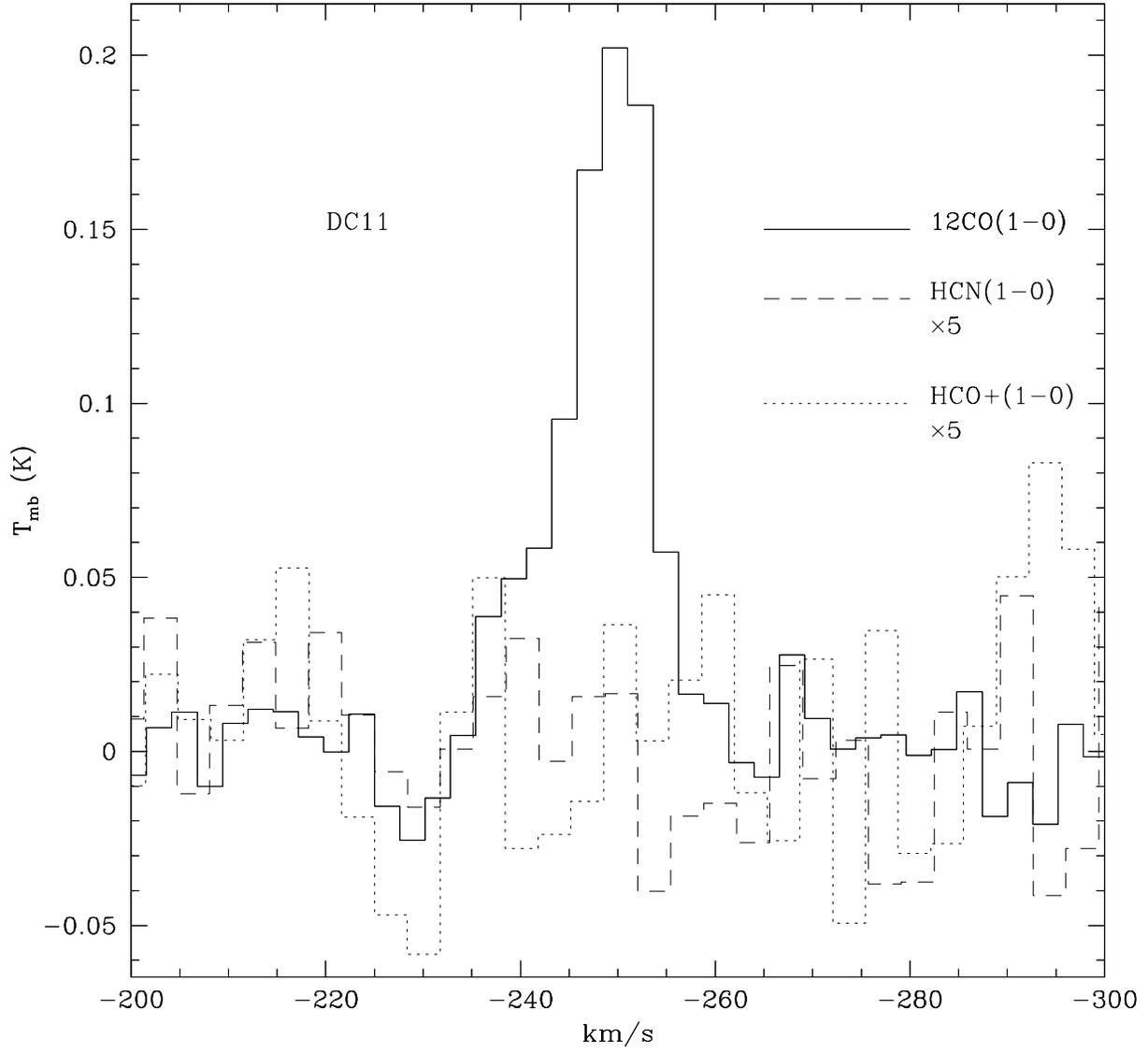

Fig. 7.— Spectra of $^{12}$CO(1-0), HCN (1-0), and HCO+ (1-0) in DC11. HCN and HCO+ spectra are scaled up for visibility.

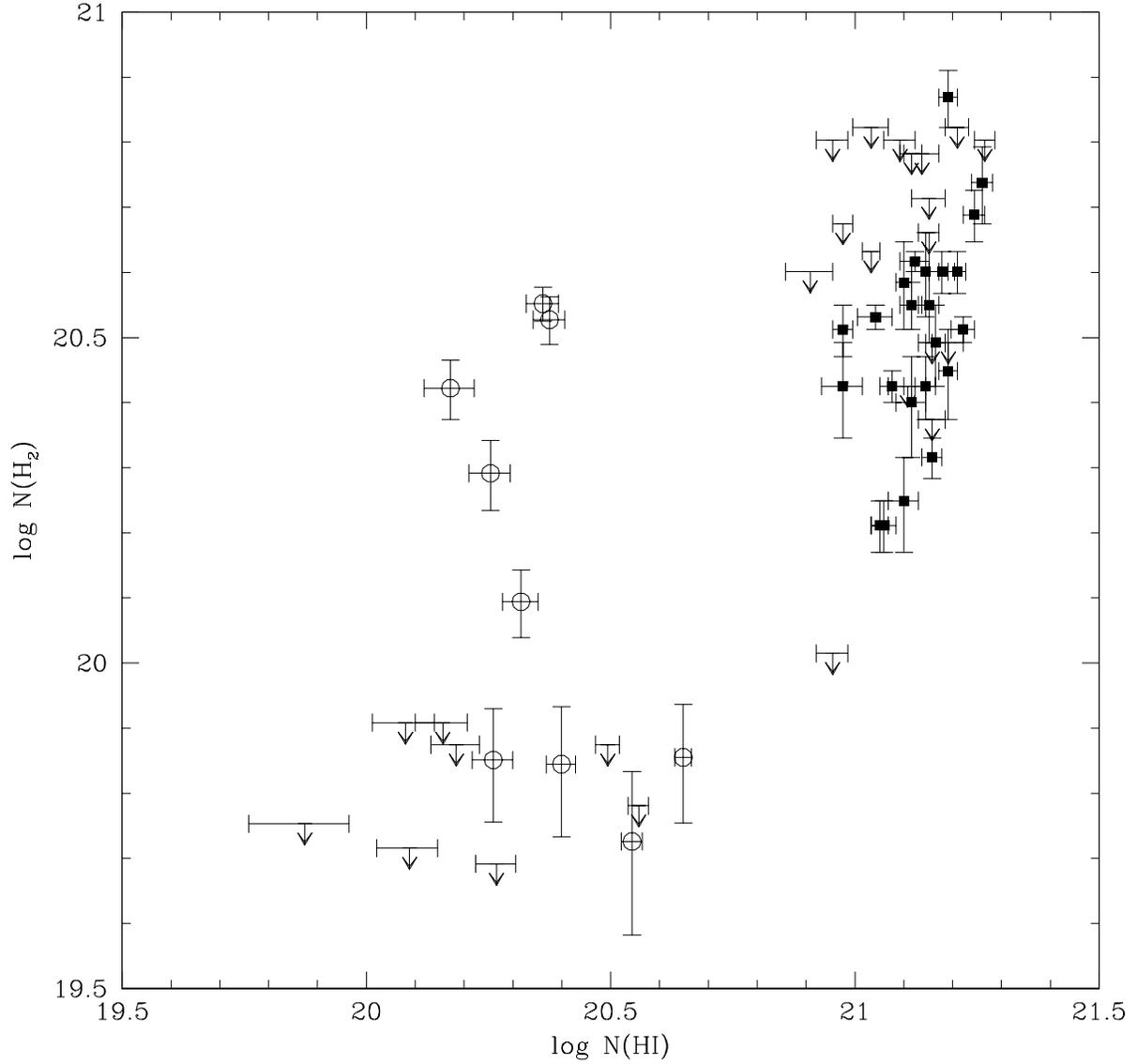

Fig. 8.— HI and H2 column densities in NGC 205 and in a spiral arm of M31 (Lada et al. 1988). Open circles are observations from the present paper. Small black squares are the results for M31, after correction for an inclination of 77° and for the different CO-to-H$_2$ conversion factor.

Table 1.   Beam Sizes and Main Beam Efficiencies

| Line | Rest Freq. GHz | FWHM $''$ | $\eta_{mb}$ | $\Delta v$ km s$^{-1}$ |
|------|------|------|------|------|
| HCN(1-0) | 88.6 | 27 | 0.82 | 3.38 |
| HCO+(1-0) | 89.2 | 27 | 0.82 | 3.36 |
| $^{13}$CO(1-0) | 110.2 | 22 | 0.74 | 2.72 |
| $^{12}$CO(1-0) | 115.3 | 21 | 0.73 | 2.60 |
| $^{13}$CO(2-1) | 220.4 | 11 | 0.48 | 1.36 |
| $^{12}$CO(2-1) | 230.5 | 10 | 0.45 | 1.30 |

Note. — Taken from Kramer & Wild (1994) and Guelin, Kramer, & Wild (1995).

Table 2.   Center Coordinates of the Observed Clouds

| Source | R.A. | Dec. |
|--------|------|------|
| | J2000.0 | |
| DC11 | 00 40 24.1 | 41 41 50.4 |
| NoDust1 | 00 40 24.0 | 41 41 20.0 |
| DC2 | 00 40 21.1 | 41 41 13.4 |
| NoDust2 | 00 40 24.5 | 41 40 50.0 |
| DC3 | 00 40 21.6 | 41 40 44.0 |
| DC12 | 00 40 25.9 | 41 40 19.4 |
| DC4 | 00 40 21.6 | 41 39 49.4 |

Table 3. $^{12}$CO(1-0) Results and H$_2$ Masses

| Source | Offsets " | rms mK | Range km s$^{-1}$ | Integ. Area K km s$^{-1}$ | Gauss area K km s$^{-1}$ | center km s$^{-1}$ | width km s$^{-1}$ | H$_2$ mass $10^4$ M$\odot$ |
|--------|-----------|--------|-------------------|---------------------------|-------------------------|-------------------|-------------------|---------------------------|
| DC11 | (0,0) | 14.8 | (-264,-234) | 2.285±0.136 | 2.158±0.120 | -249.2±0.3 | 10.0±0.7 | 5.7±0.3 |
| DC11 | (0,10) | 19.5 | (-264,-234) | 1.695±0.177 | 1.493±0.146 | -250.1±0.4 | 8.7±1.1 | 4.2±0.4 |
| DC11 | (-10,0) | 20.0 | (-264,-234) | 2.159±0.182 | 2.205±0.161 | -247.6±0.4 | 10.8±0.9 | 5.4±0.5 |
| DC11 | (10,0) | 18.9 | (-264,-234) | 0.353±0.173 | 0.511±0.117 | -250.1±0.8 | 7.1±1.8 | ⋯ |
| DC11 | (0,-10) | 17.1 | (-264,-234) | 1.254±0.155 | 0.927±0.185 | -249.8±1.0 | 8.5±2.8 | 3.1±0.4 |
| NoDust1 | (0,0) | 10.0 | (-260,-210) | 0.063±0.121 | ⋯ | ⋯ | ⋯ | ⋯ |
| DC2 | (0,0) | 8.9 | (-255,-215) | 0.796±0.095 | 0.761±0.092 | -235.6±1.2 | 19.2±2.5 | 2.0±0.2 |
| NoDust2 | (0,0) | 14.4 | (-260,-210) | -0.204±0.173 | ⋯ | ⋯ | ⋯ | ⋯ |
| DC3 | (0,0) | 13.3 | (-260,-210) | 0.058±0.160 | ⋯ | ⋯ | ⋯ | ⋯ |
| DC12 | (0,0) | 11.2 | (-235,-205) | 0.448±0.101 | 0.457±0.063 | -220.0±0.3 | 5.2±0.9 | 1.1±0.3 |
| DC12 | (0,10) | 10.0 | (-235,-205) | 0.455±0.090 | 0.376±0.052 | -219.1±0.4 | 4.8±0.7 | 1.1±0.2 |
| DC12 | (0,-10) | 11.6 | (-235,-205) | 0.210±0.105 | ⋯ | ⋯ | ⋯ | ⋯ |
| DC12 | (-10,0) | 10.5 | (-235,-205) | 0.341±0.096 | 0.325±0.072 | -217.1±0.9 | 8.2±2.2 | 0.9±0.2 |
| DC12 | (10,0) | 12.2 | (-235,-205) | 0.218±0.111 | ⋯ | ⋯ | ⋯ | ⋯ |
| DC12 | (-25,1) | 14.2 | (-235,-205) | 0.281±0.129 | ⋯ | ⋯ | ⋯ | ⋯ |
| DC12 | (-30,-9) | 10.4 | (-233,-203) | 0.459±0.095 | 0.416±0.108 | -218.0±1.5 | 13.5±5.0 | 1.1±0.2 |
| DC4 | (0,0) | 13.3 | (-260,-210) | 0.245±0.160 | ⋯ | ⋯ | ⋯ | ⋯ |

Note. — Column 1 gives the cloud name, and column 2 refers to the position within the cloud (see Figure 1). Column 4 gives the velocity range which was used to determine the integrated area in column 5. The error estimates in Tables 3 and 4 are formal (statistical) uncertainties only, and do not include the uncertainty in the temperature calibration scale (see Section 2).

Table 4.   Other Lines

| Line | Source | Offsets ″ | rms mK | Range km s⁻¹ | Integ. Area K km s⁻¹ | Gauss area K km s⁻¹ | center km s⁻¹ | width km s⁻¹ |
|------|--------|-----------|--------|--------------|----------------------|---------------------|---------------|--------------|
| $^{12}$CO(2-1) | DC11 | (0,0) | 17.2 | (-264,-234) | 1.599 ± 0.114 | 1.666 ± 0.084 | -249.5 ± 0.2 | 8.2 ± 0.5 |
| | DC11 | (0,10) | 57.0 | (-264,-234) | 1.111 ± 0.378 | 1.265 ± 0.286 | -249.7 ± 1.2 | 10.1 ± 2.1 |
| | DC11 | (-10,0) | 58.1 | (-264,-234) | 0.719 ± 0.385 | · · · | · · · | · · · |
| | DC11 | (10,0) | 58.0 | (-264,-234) | 0.114 ± 0.385 | · · · | · · · | · · · |
| | DC11 | (0,-10) | 54.7 | (-264,-234) | 0.900 ± 0.363 | · · · | · · · | · · · |
| | NoDust1 | (0,0) | 32.3 | (-260,-210) | -0.042 ± 0.286 | · · · | · · · | · · · |
| | DC2 | (0,0) | 23.5 | (-255,-215) | 0.315 ± 0.176 | 0.483 ± 0.112 | -239.2 ± 1.1 | 9.1 ± 2.1 |
| | NoDust2 | (0,0) | 46.9 | (-260,-210) | -0.132 ± 0.416 | · · · | · · · | · · · |
| | DC3 | (0,0) | 50.4 | (-260,-210) | 0.902 ± 0.425 | · · · | · · · | · · · |
| | DC12 | (0,0) | 22.5 | (-235,-205) | 0.262 ± 0.150 | · · · | · · · | · · · |
| | DC12 | (0,10) | 25.1 | (-235,-205) | 0.350 ± 0.167 | · · · | · · · | · · · |
| | DC12 | (0,-10) | 27.6 | (-235,-205) | 0.262 ± 0.178 | · · · | · · · | · · · |
| | DC12 | (-10,0) | 26.6 | (-235,-205) | 0.264 ± 0.172 | · · · | · · · | · · · |
| | DC12 | (10,0) | 30.7 | (-235,-205) | -0.224 ± 0.196 | · · · | · · · | · · · |
| | DC12 | (-25,1) | 34.6 | (-235,-205) | 0.191 ± 0.229 | · · · | · · · | · · · |
| | DC12 | (-30,-9) | 24.6 | (-233,-203) | 0.361 ± 0.158 | · · · | · · · | · · · |
| | DC4 | (0,0) | 31.9 | (-260,-210) | 0.568 ± 0.268 | · · · | · · · | · · · |
| $^{13}$CO(1-0) | DC11 | (0,0) | 3.0 | (-264,-234) | 0.244 ± 0.028 | 0.266 ± 0.024 | -248.0 ± 0.5 | 10.9 ± 1.1 |
| | DC11 | (0,10) | 10.2 | (-264,-234) | 0.189 ± 0.094 | · · · | · · · | · · · |
| | DC11 | (10,0) | 7.4 | (-264,-234) | 0.073 ± 0.069 | · · · | · · · | · · · |

| Line | Source | Offsets " | rms mK | Range km s$^{-1}$ | Integ. Area K km s$^{-1}$ | Gauss area K km s$^{-1}$ | center km s$^{-1}$ | width km s$^{-1}$ |
|---|---|---|---|---|---|---|---|---|
| 13CO(2-1) | DC11 | (0,0) | 12.1 | (-264,-234) | 0.128 ± 0.082 | ⋯ | ⋯ | ⋯ |
| | DC11 | (0,10) | 35.3 | (-264,-234) | -0.151 ± 0.239 | ⋯ | ⋯ | ⋯ |
| | DC11 | (10,0) | 26.0 | (-264,-234) | -0.113 ± 0.176 | ⋯ | ⋯ | ⋯ |
| HCN(1-0) | DC11 | (0,0) | 5.4 | (-264,-234) | -0.015 ± 0.057 | ⋯ | ⋯ | ⋯ |
| HCO+(1-0) | DC11 | (0,0) | 7.5 | (-264,-234) | 0.060 ± 0.077 | ⋯ | ⋯ | ⋯ |
| 12CO(2-1)[a] | DC11 | (0,0) | 18.6 | (-264,-234) | 1.041±0.123 | 1.165±0.098 | -249.1±0.4 | 9.9±0.9 |
| | DC12 | (0,0) | 12.1 | (-235,-205) | 0.209±0.081 | 0.299±0.084 | -224.3±2.5 | 16.1±5.1 |

[a]Convolved to 21″ resolution.

Note. — Columns are similar to those in Table 3. The error estimates in Tables 3 and 4 are formal (statistical) uncertainties only, and do not include the uncertainty in the temperature calibration scale.

Table 5.   Line Ratios

| Lines | Source | Offsets $''$ | Moment ratio | Gauss. fit ratio |
|---|---|---|---|---|
| $^{12}CO(1\text{-}0)/^{12}CO(2\text{-}1)$ | DC11 | ( 0, 0) | $2.20 \pm 0.29$ | $1.85 \pm 0.19$ |
| $^{12}CO(1\text{-}0)/^{12}CO(2\text{-}1)$ | DC12 | ( 0, 0) | $> 1.83$ | $1.53 \pm 0.48$ |
| $^{12}CO(1\text{-}0)/^{13}CO(1\text{-}0)$ | DC11 | ( 0, 0) | $9.35 \pm 1.22$ | $8.12 \pm 0.86$ |
| $^{12}CO(2\text{-}1)/^{13}CO(2\text{-}1)$ | DC11 | ( 0, 0) | $> 6.51$ | $\cdots$ |
| $^{12}12CO(1\text{-}0)/HCN(1\text{-}0)$ | DC11 | ( 0, 0) | $> 13.5$ | $\cdots$ |
| $^{12}12CO(1\text{-}0)/HCO+(1\text{-}0)$ | DC11 | ( 0, 0) | $> 9.83$ | $\cdots$ |

Note. — Ratios of (1-0)/(2-1) lines are made after convolving the 2-1 data to the same resolution as the 1-0 data. The uncertainty values in the table (10–13% for $^{12}CO$ (1-0)/(2-1) and $^{12}CO(1\text{-}0)/^{13}CO(1\text{-}0)$) are $1\sigma$ estimates for the ratio of two independent random variables (Bevington & Robinson 1992), and they do not include calibration uncertainties. The uncertainties from the calibration scale are also about 10–15% for each line (Section 2). Thus, the true uncertainties in the line ratios are about $\sqrt{2}$ larger than the formal values in the table.

Table 6.   Molecular Line Ratios in NGC 205 and in Other Environments

| Ratio | NGC 205 | Other Ellipticals | Milky Way Disk or Other Spiral Disk | Bulges or Nuclei of Nearby Spirals | References |
|---|---|---|---|---|---|
| $^{12}$CO 2-1/1-0 | 0.5±0.1 | 0.7 − 1.5[a] | 0.6 − 1.0 | 0.2 − 0.4[b] | this paper; WCH95; SST93, SHH97; A95 |
| $^{12}$CO 1-0/$^{13}$CO 1-0 | 8.7±1.5 | 6 − 16 | 5 − 10 | 5 − 17 | this paper; R97; SST93; SI91 |
| $^{12}$CO 1-0/HCN 1-0 | > 13 | 8.7±1.7[c] | ∼40 and up | ≲ 12 | this paper; I92; HB97 |
| $^{12}$CO 1-0/HCO+ 1-0 | > 10 | 8.3±1.7[c] | 25 and up | · · · | this paper; I92; L95 |

[a]These values are upper limits, by factors of 1 to 4, because they have not been corrected for different beam sizes (WCH95).

[b]In the bulge of M31 (Allen et al. 1995).

[c]In the circumnuclear disk of the peculiar elliptical NGC 5128 = Centaurus A (Israel 1992).

Note. — References: WCH95 = Wiklind, Combes, & Henkel (1995). SST93 = Sanders et al. (1993). SHH97 = Sakamoto et al. (1997). A95 = Allen et al. (1995). R97 = Rupen (1997), which is a compilation of results from the literature. SI91 = Sage & Isbell (1991). I92 = Israel (1992). HB97 = Helfer & Blitz (1997a, 1997b). L95 = Liszt (1995).

Table 7.  Atomic and Molecular Column Densities

| Source | Offsets $''$ | log N(HI) atoms cm$^{-2}$ | log N(H$_2$) mol cm$^{-2}$ | log N(H) H cm$^{-2}$ |
|---|---|---|---|---|
| DC11 | ( 0, 0) | $20.361 \pm 0.032$ | $20.552 \pm 0.025$ | $20.974 \pm 0.021$ |
| DC11 | ( 0, 10) | $20.172 \pm 0.048$ | $20.422 \pm 0.043$ | $20.831 \pm 0.036$ |
| DC11 | (-10, 0) | $20.375 \pm 0.031$ | $20.527 \pm 0.035$ | $20.959 \pm 0.027$ |
| DC11 | ( 10, 0) | $20.080 \pm 0.059$ | $< 19.908$ | (20.080, 20.451) |
| DC11 | ( 0,-10) | $20.254 \pm 0.040$ | $20.291 \pm 0.051$ | $20.756 \pm 0.037$ |
| NoDust1 | ( 0, 0) | $19.874 \pm 0.091$ | $< 19.753$ | (19.874, 20.274) |
| DC2 | ( 0, 0) | $20.317 \pm 0.035$ | $20.094 \pm 0.049$ | $20.659 \pm 0.032$ |
| NoDust2 | ( 0, 0) | $20.157 \pm 0.050$ | $< 19.908$ | (20.157, 20.485) |
| DC3 | ( 0, 0) | $20.184 \pm 0.047$ | $< 19.874$ | (20.184, 20.481) |
| DC12 | ( 0, 0) | $20.399 \pm 0.029$ | $19.844 \pm 0.088$ | $20.592 \pm 0.038$ |
| DC12 | ( 0, 10) | $20.260 \pm 0.040$ | $19.851 \pm 0.078$ | $20.510 \pm 0.042$ |
| DC12 | ( 0,-10) | $20.266 \pm 0.039$ | $< 19.691$ | (20.266, 20.452) |
| DC12 | (-10, 0) | $20.544 \pm 0.021$ | $19.726 \pm 0.108$ | $20.659 \pm 0.032$ |
| DC12 | ( 10, 0) | $20.088 \pm 0.058$ | $< 19.716$ | (20.088, 20.355) |
| DC12 | (-25, 1) | $20.558 \pm 0.020$ | $< 19.781$ | (20.558, 20.683) |
| DC12 | (-30, -9) | $20.649 \pm 0.017$ | $19.855 \pm 0.082$ | $20.770 \pm 0.025$ |
| DC4 | ( 0, 0) | $20.494 \pm 0.024$ | $< 19.874$ | (20.494, 20.665) |

Note. —  HI column densities are extracted from the data of Young & Lo (1997), which was reprocessed to have the same beam size as the CO observations (a circular beam of 21$''$ FWHM). The total gas column density N(H) is a weighted sum N(H) = N(HI) + 2 N(H$_2$) nuclei cm$^{-2}$. Where CO has not been detected, the column density N(H$_2$) is a $3\sigma$ upper limit as discussed in the text. In those cases the total column density N(H) is given as a range from N(HI) to N(HI) plus twice the N(H$_2$) limit.

Table 8.   Estimates of $N_H/A_V$ in NGC 205

| | Cloud | |
| --- | --- | --- |
| | DC2 | DC11 |
| $N_H$, cm$^{-2}$ | $4.6\times10^{20}$ | $9.4\times10^{20}$ |
| Apparent $A_V$, mag | 0.47 | 0.21 |
| Corrected $A_V$, mag | 2 | 0.42 |
| $N_H/A_V$ | $2.3\times10^{20}$ | $2.2\times10^{21}$ |

Note. — Gas column densities in line 1 are taken from table 7.